\documentclass[conference]{IEEEtran}
\ifCLASSINFOpdf

\else

\fi

\usepackage{amsmath}
\usepackage{amssymb}
\usepackage[dvips]{graphicx}
\usepackage{epsfig}
\usepackage[compress]{cite}
\usepackage{caption,keywords}
\usepackage{color}
\usepackage{lipsum}
\usepackage{enumerate}
\usepackage{algorithm,algpseudocode}

\newtheorem{prop}{Proposition}

\newcommand{\am}[1]{\textcolor{black}{#1}\marginpar{\footnotesize{\textcolor{blue}{   }}}}

\newcommand{\mar}{\textcolor{black}}
\begin{document}

\title{Design of a Cognitive VLC Network with Illumination and Handover Requirements}
    
\author{\IEEEauthorblockN{Marwan Hammouda\IEEEauthorrefmark{1}, J\"{u}rgen Peissig\IEEEauthorrefmark{1}, and
Anna Maria Vegni\IEEEauthorrefmark{2}}\\
\IEEEauthorblockA{
\IEEEauthorrefmark{1}Institute of Communications Technology\\Leibniz Universit\"{a}t Hannover, Hannover, Germany \\ Email: \{marwan.hammouda, peissig\}@ikt.uni-hannover.de\\
\IEEEauthorrefmark{2}Department of Engineering\\
University of Roma Tre, Rome, Italy\\
Email: annamaria.vegni@uniroma3.it}}

\maketitle

\begin{abstract}
In this paper, we consider a cognitive  indoor visible light communications (VLC) system, 
comprised of multiple  access points  serving  primary and secondary users through 
the orthogonal frequency division multiple access  method.  
A cognitive lighting cell is divided  into two non-overlapping regions that distinguish the  primary and secondary 
users based on the region they are located in. Under the assumption of equal-power allocation among subcarriers, each region is defined in terms of its physical area and the number of allocated subcarriers within that region. In this paper, we provide the lighting cell design with cognitive constraints that guarantee fulfilling certain illumination, user mobility, and handover requirements in each cell. We further argue that, under some conditions, a careful assignment of the subcarriers in each region can mitigate the co-channel interference in the overlapping areas  of adjacent cells. Numerical results depict the influence of different system parameters, such as user density, on defining both regions. Finally, a realistic example is implemented to assess the performance of the proposed scheme via  Monte Carlo simulations.
\end{abstract} 

\section{Introduction}
Cognitive radio (CR)~\cite{haykin2005cognitive} and visible light communications~\cite{elgala2011indoor} have been gaining an increased attention as promising solutions for the over-crowded radio frequency (RF) spectrum problem, known as \textit{spectrum crunch problem}~\cite{karamchandani2014agile}. While the CR approach aims at improving the utilization of the existing RF spectrum by introducing the concepts of primary (licensed) and secondary (unlicensed) users, 
the VLC technology suggests using the visible bands of the electromagnetic spectrum. 


However, the VLC technology comes with other restrictions and challenges that should be carefully handled. 
For instance, the main lighting functionality of the LED-based access points  (expressed in terms of illumination requirements) 
should be considered when designing the VLC networks. In this regard, the authors in~\cite{stefan2013analysis} demonstrated the influence of the illumination requirements on the optimal 
placement of the LED-based access points (APs). Furthermore, the authors in~\cite{Standard,grubor2008broadband} proposed a 
``cell zooming'' method in order to achieve constant illumination levels over the entire space by either altering the 
transmitted power or physically changing the radiation pattern. 

In addition to the lighting needs, each VLC cell can cover a small area of square meters, and hence typical indoor scenarios are equipped with multiple lighting sources to cover the entire area. This further makes the network planning more challenging when dealing with mobile users, since the connectivity switching process from one AP to another (known as handover) is expected to occur more often. While each handover process requires 
extra signaling overhead, the user connectivity also becomes an issue during the switching process. 

The handover process in VLC systems has been investigated by many researchers~\cite{vegni2012handover,dinc2015soft,liang2015novel,rahaim2013sinr,bui2014leds}. However, these studies did not consider the signaling overhead needed by the switching process, or the illumination constraints. Different than that, 
the authors in~\cite{pergoloni2016optimized} investigated the optimal footprinting of a VLC cell that maximizes the user average rate, while considering mobility and handover overhead under the assumption of using the time division multiple access (TDMA) as the medium access technique. However, the authors neglected the other aspects of user connectivity and lighting requirements. 

In this paper, we consider an indoor VLC network, in which multiple LED-based APs serve mobile users by employing the orthogonal frequency division multiple access (OFDMA) scheme. Motivated by the CR concept, we  assume that each lighting cell can be  divided into two  regions that distinguish mobile users  into primary and secondary users based on their 
location information. Under the assumption of equal power allocation among all subcarriers, we investigate the effects of fulfilling illumination, mobility, and handover requirements on defining each region in terms of physical area and the number of allocated subcarriers. Consequently, we formulate a design criteria that ensures fulfilling the mentioned constraints. Finally, we consider a realistic indoor scenario and investigate the performance of the proposed method using  Monte Carlo simulations.
 
 This paper is organized as follows. Section~\ref{sec:System_Model} introduces our main concept of cognitive VLC networks. Section~\ref{sec:System_Model} presents our proposed resource allocation scheme by addressing the requirements due to illumination and handover. A realistic scenario is implemented in Section~\ref{sec:example} and simulation results are expressed in terms of the average failure rate. Finally, conclusions are drawn at the end of this paper. 
 
\begin{figure}[t]
\begin{center}
\includegraphics[width=0.3\textwidth]{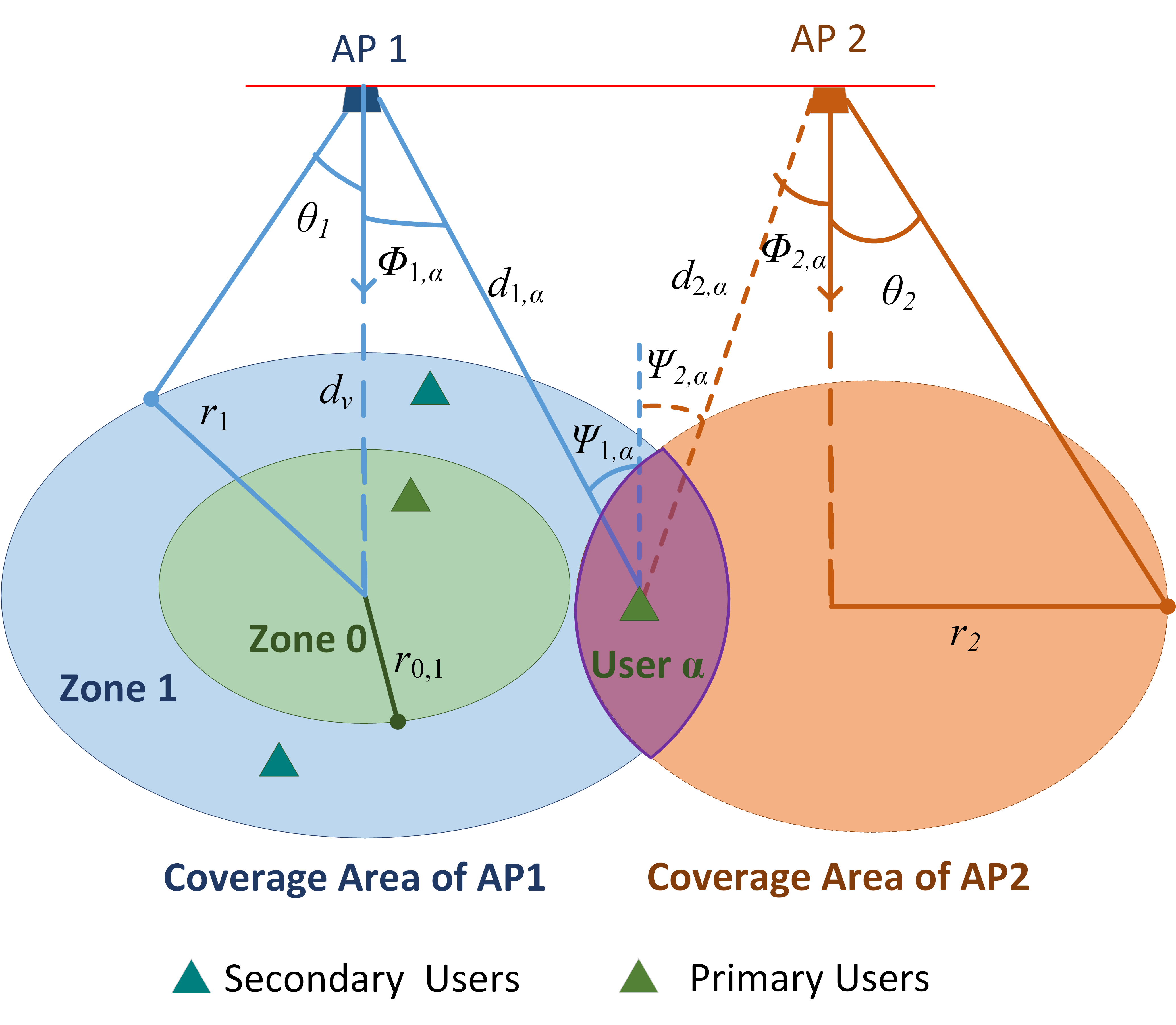}
\caption{General system model of a cognitive VLC scenario, comprised of two LED-based access points (\textit{i.e.}, AP1 and AP2) and multiple mobile users.}\label{Channel_Model}
\end{center}
\end{figure}

\section{System Model}
\label{sec:System_Model}
Let us consider an indoor communication scenario  consisting of $K$ LED-based APs 
and $M$ mobile users. We regard a downlink transmission scenario and assume that each AP employs 
OFDMA scheme to serve multiple users within its coverage area. 
To simplify,  we assume that each AP produces an ideal cone of light, \textit{i.e.}, its entire light output is projected as 
a circular lighting field with a hard boundary, centered at the AP location. 

Although VLC links normally include both line-of-sight (LoS) and non-LoS (diffuse) components, 
without loss of generality we can focus on  LoS links only~\cite{komine2004fundamental}. A simple illustration of the considered scenario is shown in \figurename~\ref{Channel_Model} for $K = 2$. The coverage area of each AP is divided into two regions, namely Zone $0$ and 
Zone $1$. The users located in Zone $0$ are assumed to be primary users, whereas the users of Zone $1$ 
are referred as secondary ones. Without loss of generality, we assume that primary (secondary) users are uniformly distributed in 
Zone $0$ (Zone $1$). \mar{In this paper, we assume that each AP has a full knowledge about the locations of all users (primary or secondary) within its coverage area~\footnote{In this paper we omit how to get the user location information.}.} 

Different than the existing literature studies, which aim at optimizing the coverage area of a VLC cell, we define 
Zone $0$ and Zone $1$ in terms of both the geographical area and the amount of resources allocated for each region. 
To that end, the radius of Zone $0$ related to the $k$-th AP is denoted as $r_{0,k}$~[m], whereas the radius of the entire coverage area is denoted as $r_k$~[m] for $k = 1,\dots,K$. It follows that Zone $1$ is defined as a two-dimensional ring whose width is $r_{1,k} = r_k - r_{0,k}$~[m]. 
Note that in general cases when no certain constraints are considered, we have $r_{0,k} \leq r_k$, and hence $r_{1,k} \geq 0$~\footnote{Throughout this paper, all distances have the unit of meters [m].}. 

Let $P_{\text{cell},k}$, $B_{\text{cell},k}$, and $N_{\text{cell},k}$ be, respectively, the total optical power, the total bandwidth, and the total number of subcarriers of the $k$-th AP. Subsequently, we assume that $P_{z,k}$, $B_{z,k}$, and $N_{z,k}$ are the corresponding resources allocated for Zone $z$ for $z = \{0,1\}$. By assuming that all cell resources are used for data transmission, the following equations hold: 
\begin{equation*}
P_{0,k} + P_{1,k} = P_{\text{cell},k},
\end{equation*}
\begin{equation*}
B_{0,k} + B_{1,k} = B_{\text{cell},k},
\end{equation*}
and
\begin{equation*}
N_{0,k} + N_{1,k} = N_{\text{cell},k}.
\end{equation*}
We  assume that all APs have the same optical power, bandwidth, and the number of subcarriers, \textit{i.e.,} $P_{\text{cell},k} = P_{\text{cell}}$, $B_{\text{cell},k} = B_{\text{cell}}$, and $N_{\text{cell},k} = N_{\text{cell}}$, respectively, for $k = 1,\dots,K$. Also, we assume that all subcarriers in each AP are allocated the same bandwidth and power. Let $B_{j,k}$ and $P_{j,k}$ be the allocated bandwidth and power at the $j$-th subcarrier and 
the $k$-th AP, respectively. Then, we have $B_{j,k} = B_{\text{cell}}/N_{\text{cell}}$ and $P_{j,k} = P_{\text{cell}}/N_{\text{cell}}$ for $k = 1,\dots,K$ and $j = 1,\dots, N_{\text{cell}}$. 
Consequently, we have $P_{z,k} = N_{z,k}P_{\text{cell}}/N_{\text{cell}}$ and $B_{z,k} = N_{z,k}B_{\text{cell}}/N_{\text{cell}}$ for $z = \{0,1\}$. From these definitions and assumptions, we note that Zone $z$ (with $z = \{0,1\}$) of the $k$-th AP can be completely characterized by finding its radius $r_{z,k}$, and the allocated number of subcarriers $N_{z,k}$. 

Through this paper, all APs reuse the same bandwidth, and hence the users located 
in an overlapping area of two or more adjacent APs might experience a co-channel interference (CCI). However, we 
assume that the subcarriers allocated in Zone $1$ have different frequency bands in the adjacent cells, 
\am{and that the overlapping area of any two adjacent cells occurs only over Zone $1$ in both cells (see purple area in \figurename~\ref{Channel_Model}). } Consequently, the CCI in 
the overlapping areas can be mitigated. This approach can be realized when each AP shares its information with all adjacent APs, 
and hence they can perform a collaborative resource management, or alternatively,  when the VLC network has a central AP, which is responsible for resource allocation in each cell based on a prior knowledge about the entire network. Accordingly, in this paper we assume that the coverage area of each cell is free of the CCI.

Let a user, denoted as $\alpha$, be located within the coverage area of the $k$-th AP, \textit{e.g.}, AP1 as depicted in \figurename~\ref{Channel_Model} for $k = 1$. Assuming that the AP follows the Lambertian radiation pattern, then the channel gain is given as~\cite{barry1993simulation}:
\begin{equation}
\label{VLC_gain_1}
h_{k,\alpha}  = \frac{(m_k+1)A_d}{2 \pi d_{k,\alpha}^2} \cos^{m_k}(\phi_{k,\alpha}) g(\psi_{k,\alpha}) \cos(\psi_{k,\alpha}),
\end{equation}
where $g(\psi_{k,\alpha})$ is the optical concentrator gain, $A_d$ is the photodiode (PD) physical area, $d_{k,\alpha}$ is the distance between the AP and the user, $\phi_{k,\alpha}$ and $\psi_{k,\alpha}$ are, respectively, the irradiance and incidence angles, and $m_k = -1/\log_2(\cos(\theta_k))$ is the Lambertian index, where $\theta_k$ is the LED half intensity viewing angle of the $k$-th AP. In this paper, we assume that all APs are directed downwards  and all users are directed upwards. Then, we have $\cos(\phi_{k,\alpha}) = \cos(\psi_{k,\alpha}) = {d_v}/{d_{k,\alpha}}$,  where $d_v$ is the normal distance between the transmitter and receiver planes. It follows that the per-subcarrier signal-to-noise ratio (SNR) at the user $\alpha$ can be expressed as 
\begin{equation}
\label{SINR_general}
\text{SNR}_{k,\alpha}^{\text{sub}} = \frac{(\gamma_{\alpha} P_{\text{sub}} h_{k,\alpha})^2}{\varsigma N_n B_{\text{sub}}},
\end{equation}
where $\gamma_{\alpha}$ is the receiver optical-to-electrical conversion efficiency,  $N_n$ [A$^{2}$/Hz] is the 
noise power spectral density, and $\varsigma$ is the ratio between the average optical power and the average electrical power of the transmitted signal. 

Finally, from \eqref{SINR_general},  the per-subcarrier achievable rate (capacity lower bound) [bits/s] is given as \cite{chaaban2016fundamental}
\begin{equation}
\label{Rate_general}
R_{k,\alpha}^{\text{sub}} = \frac{B_{\text{cell}}}{2} \log_2(1+c^2 \text{SNR}_{k,\alpha}^{\text{sub}}).
\end{equation}
for some constant $c$ \footnote{For instance, $c = \sqrt{e/2 \pi} = 0.93$ when the transmitted light intensity is exponentially distributed.}. For simplicity and without loss of generality, we set $c = 1$ in this paper. 
\begin{table}
\caption{Parameters used in the simulation results.}
\begin{center}
\begin{tabular}{c|c}\hline
{{\bf Parameters}} & {\bf Values}\\
\hline
$B_{\text{cell}}$ & $20$~MHz\\ 
 $P_{\text{cell}} $ &  $9$~W\\
 $d_v$ & $3.5$~m\\ 
 $\psi_C$  & $ 90^{\circ}$ \\
 $A_d$ & $ 1 \,\text{cm}^2$\\ 
 $\gamma$ &  $0.53$ A/W\\ 
 $\varsigma$ & $1$\\
 $N_n$ & $10^{-21}$~$\text{A}^2/$Hz\\
\hline
\end{tabular}
\end{center}
\label{tab_1}
\end{table}%

\section{System Analysis}\label{sec:analysis}
In this section, we explore the effects of different system settings and requirements on defining Zone $0$ and Zone $1$. In particular, we address the illumination and handover needs. For presentation purposes, and without loss of generality, in this section we carry out the analysis for AP1  shown in \figurename~\ref{Channel_Model}.

We initially consider the case when the system aims at maximizing the area spectral efficiency (ASE), defined as the average data rate per unit
bandwidth per unit area supported within a lighting cell,  without imposing any certain constraints. 
%
Let $\bar{R}_{(z, k=1)}^{\text{sub}}$ be the average rate achieved by a single subcarrier in Zone $z$ for $z = \{0,1\}$. We can readily observe that $\bar{R}_{(0,k=1)}^{\text{sub}} > \bar{R}_{(1,k=1)}^{\text{sub}}$. Recalling that the entire coverage area of AP1 is assumed to be CCI-free, and under the assumption of uniform user distribution in each zone, we initially provide the following preposition~\footnote{The proof is omitted due to space limitation.} that characterizes the average rates $\bar{R}_{(z, k=1)}^{\text{sub}}$. \\

\begin{prop}
For AP1 with LoS links and CCI-free coverage area, if users in Zone $z$ with $z = \{0,1\}$ are uniformly distributed, then the average rate achieved by a single subcarrier can be expressed as: 
{\small
\begin{equation*}
\label{avg_rate}
\begin{aligned}
& \bar{R}_{(z, k=1)}^{\text{sub}} =
 \frac{d_{\text{max}} [\ln(\rho \kappa_{\text{min}})+m_1 + 3] - d_{\text{min}} [\ln(\rho\kappa_{\text{max}})+m_1+3]}{2 \ln(2) (r_{\text{max}}^2 - r_{\text{min}}^2)},
\end{aligned}
\end{equation*}}
where $r_{\text{min}} = 0$ and $r_{\text{max}} = r_{0,1}$ for Zone $0$, whereas $r_{\text{min}} = r_{0,1}$ and $r_{\text{max}} = r_1$ for Zone $1$. In addition, $d_{\text{max}} = r_{\text{max}}^2 + d_v^2$, $d_{\text{min}} = r_{\text{min}}^2 + d_v^2$ , $\rho = \frac{P_{\text{sub}}^2 \gamma^2}{N_n B_{\text{sub}}}$, $\kappa_{\text{min}} = \frac{(A_d g(\psi_1) (m+1)d_v^{m_1+1})^2}{(2\pi)^2 d_{\text{max}}^{m_1+3}}$, and $\kappa_{\text{max}} = \frac{(A_d g(\psi_1) (m+1)d_v^{m_1+1})^2}{(2\pi)^2 d_{\text{min}}^{m_1+3}}$.
\end{prop}
\vspace{0.2cm}
Note that both $\bar{R}_{0,1}^{\text{sub}}$ and $\bar{R}_{1,1}^{\text{sub}}$ have higher values at smaller values of $r_{0,1}$ such that they are both maximized when $r_{0,1} = 0$. Now, let the ASE of AP1 be denoted as $\eta_1$ $[\text{bits}/\text{s}/\text{m}^2]$. Noting that the total average rate in Zone $z$ (with $z=\{0,1\}$) is equal to $\bar{R}_{(z, k =1)} = N_{(z, k=1)} \bar{R}_{(z, k= 1)}^{\text{sub}}$, then we have:
\begin{align}
\label{ASE_proposed}
\eta_1 = & \frac{N_{0,1} \bar{R}_{0,1}^{\text{sub}} + (N_{\text{cell}} - N_{0,1}) \bar{R}_{1,1}^{\text{sub}}}{\pi B_{\text{cell}} r_1^2} \notag \\
= & \frac{N_{0,1}}{N_{\text{cell}}} \frac{\bar{R}_{0,1}^{\text{sub}} - \bar{R}_{1,1}^{\text{sub}}}{\pi B_{\text{cell}r_1^2}} + \frac{\bar{R}_{1,1}^{\text{sub}}}{\pi B_{\text{cell}r_1^2}}.
\end{align}
\begin{figure}[t]
\begin{center}
\includegraphics[width=0.35\textwidth]{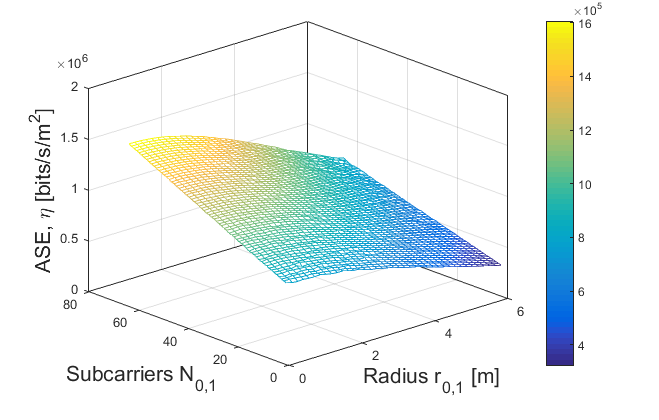}
\caption{Area spectral efficiency of AP1 as a function of both $r_{0,1}$ and $N_{0,1}$ when $\theta_1 = 60^{\circ}$ ($r_1 \approx 6 $ m) and $N_{\text{cell}} = 64$.} \label{ASE_3D}
\end{center}
\end{figure}
In \figurename~\ref{ASE_3D}, we depict $\eta_1$ as a function of both $r_{0,1}$ and $N_{0,1}$ when $\theta_1 = 60^{\circ}$, while other parameters are as given in Table~\ref{tab_1}. We  observe that the ASE has the maximum value (\textit{i.e.}, $1.47$~Mbit/s/m$^2$) when allocating the total available subcarriers in Zone $0$, \textit{i.e.}, when $N_{0,1} = N_{\text{cell}}$, while setting its radius to the minimum value,  \textit{i.e.},  $r_{0,1} = 0$. 
However, such settings are not practically feasible. 
Indeed, the radius of Zone $0$ 
should be large enough to ensure serving a sufficient number of primary users. Assuming  
$\epsilon$ [user/m$^2$] as the user density in the indoor environment, the number of primary users that can be served by AP1 is $U_{0,1} = \pi \epsilon r_{0,1}^2$. To give an example, if a system targets serving $2$ primary users, then we have $r_{0,1} \geq 1.78$~m if $\epsilon = 0.2$ user/m$^2$. On the other hand, leaving Zone $1$ without any subcarriers is not preferred in many practical scenarios. This is specially true when primary users are mobile, and hence subcarriers in Zone $1$ can be used to support the handover process when a primary user leaves Zone $0$. Therefore, maximizing the ASE, while considering other system requirements, is a critical objective. 

\subsection{Illumination Requirements}
Since LED-based APs are originally used for lighting purposes, the need for a sufficient amount of light over the receiving plane should be considered in the system design. For instance, the  European lighting standard~\cite{Standard} defines different brightness levels that should be satisfied in various indoor scenarios, \textit{e.g.,} offices, corridors, factories, or training rooms, which highly depend on the running activities in each scenario. 

In this respect, illuminance is the most commonly used factor that characterizes the brightness factor at a given location. Formally, the illuminance level [lx] at a horizontal distance $r_{0,k}$ from the center of the $k$-th AP, \textit{i.e.}, at the edge of Zone $0$ related to that AP, can be calculated 
as~\cite{grubor2008broadband}:
\begin{equation}
\label{illumination_1}
E = I_0 \frac{d_v^{m_k + 1}}{(r_{0,k}^2 + d_v^2)^{\frac{m_k + 3}{2}}}, 
\end{equation}
where $I_0$ is the maximum luminous intensity [cd]. In this paper, we target a brightness span of $[E_{\text{min}},E_{\text{max}}]$ lx within Zone 0, such that the brightness level at the zone edge fulfills the minimum level of $E_{\text{min}}$, while it does not exceed $E_{\text{max}}$ at the cell center for the eye safety concerns. Recalling that AP1 is assumed to have a CCI-free coverage area, then we have the following constraints by solving \eqref{illumination_1}: 
\begin{equation}
\label{illumination_2}
r_{0,k} \leq \bigg[\bigg(\frac{I_0 d_v^{m_k+1}}{E_{\text{min}}}\bigg)^{\frac{2}{m_k + 3}} - d_v^2\bigg]^{\frac{1}{2}},
\end{equation}
and
\begin{equation}
E_{\text{min}} d_v^2 \leq I_0 \leq E_{\text{max}} d_v^2,
\end{equation}
respectively, on the radius $r_{0,k}$ and the maximal luminous intensity $I_0$. 

Now, let the illumination level at the cell center equals $E_{\text{max}}$, then $I_0 = E_{\text{max}} d_v^2$, 
and the limit on $r_{0,k}$ in \eqref{illumination_2} can be re-expressed as  
\begin{equation}
\label{illumination_3}
r_{0,k}^2 \leq d_v^2 \bigg[\bigg(\frac{E_{\text{max}}}{E_{\text{min}}}\bigg)^{\frac{2}{m_k + 3}} - 1\bigg]  := \Lambda_{0,1}^2, 
\end{equation}
which is a function of the ratio between the maximum and the minimum illumination levels in Zone $0$. As an example, let us set the limits to $[E_{\text{min}},E_{\text{max}}] = [200,800]$ lx, 
as considered in~\cite{grubor2008broadband}.  Then, we have 
\begin{equation}
\label{illumination_4}
r_{0,k} \leq d_v  \bigg(16^{\frac{1}{m_k + 3}} - 1\bigg)^{\frac{1}{2}}.
\end{equation}

To show the impact of imposing the above-mentioned illumination requirements on defining Zone $0$ of the $k$-th AP, let $\theta_k = 60^{\circ}$ and $d_v = 3.5$ m. Then, we have $r_{0,k} \leq r_{k} = d_v \tan(\theta_1) \approx 6$ m when no illumination requirements are considered, whereas  $r_{0,k} \leq 3.5$~m according to \eqref{illumination_4}. We finally remark that the effect of the illumination requirements is neglected when the right-hand term in (\ref{illumination_3}) is greater than the cell radius, \textit{i.e.,} when $[(E_{\text{max}}/E_{\text{min}})^{\frac{2}{m_k + 3}} - 1]^{\frac{1}{2}}>\tan(\theta_1)$. This condition is satisfied when $\theta_1 \leq 37^{\circ}$ for the ratio $E_{\text{max}}/E_{\text{min}} = 4$. 

\subsection{Mobility and Handover Requirements}
Due to users mobility, a switching process from one AP
to another might occur. This process is commonly referred as
handover, and it normally requires an extra overhead as control signaling. In addition, a seamless handover mechanism is preferred in order to maintain service connectivity without interruption while a user moves from one cell to another. 
According to  cell design, handovers are expected to happen more frequently in smaller cells, and hence the design of Zone $0$ and Zone $1$ is a critical factor for handover initialization.


In this paper, we consider Zone $0$ of each cell as the main region, which is allocated the majority of transmission resources, \textit{e.g.,} subcarriers and power. 
%
\am{Considering \figurename~\ref{Channel_Model}, the overlapping area of  two adjacent cells is served 
by AP1 and AP2. The user $\alpha$ is connected to AP2 and moves towards Zone $0$, served by AP1 only.
In this case  a  handover process from Zone $1$ to Zone $0$ occurs, and therefore, Zone $1$ is considered as 
the handover region, and it is allocated the sufficient resources to execute the switching process for all secondary users~\footnote{The case of handover from Zone $0$ to Zone $1$ is similar.}.} In addition to the handover overhead requirements, we assume that Zone $1$ is allocated additional resources to support mobile users during the switching event, and hence ensuring a seamless process. Recall that the allocation of additional resources in Zone $1$ has the advantage of suppressing 
the CCI in the overlapping areas of adjacent cells.
Note that the  handover requirements are conflicting with the objective of maximizing the ASE over the coverage area, as shown in \figurename~\ref{ASE_3D}. In the following, we explore the effects of different system parameters on defining Zone $0$ and Zone $1$, while taking the handover requirements into account.

To quantify the aforementioned assumptions,  let $U_{z,1}$ be the number of users located in Zone $z$ related to AP1 for $z = \{0,1\}$, and $\epsilon$ be the user density [user/m$^2$] within the entire indoor environment. Then, we have $U_{0,1} = \pi \epsilon r_{0,1}^2$ and $U_{1,1} = \pi \epsilon (r_1^2 -r_{0,1}^2)$. Furthermore, the bandwidth required to handle one handover process is denoted as $B_{\text{HO}}$ [bits], and the percentage of the primary users that leave Zone $0$ is denoted as $\beta$.  Herein, we consider a special case and assume that each primary user, either when being located in Zone $0$ or when moving to Zone $1$, is allocated a single subcarrier. Note that this limit case is applied when a system aims at maximizing the number of primary users, regardless of their data types. Based on these assumptions, we can formulate the following limit:
\begin{equation}
\pi \epsilon r_{0,1}^2 + \beta \pi \epsilon r_{0,1}^2 + \pi \epsilon (r_1^2 -r_{0,1}^2) \frac{B_{\text{HO}}}{B_{\text{sub}}} \leq N_{\text{cell}}, \label{lim_1}
\end{equation}
which, after some manipulations, results in the following design criteria:
\begin{align}
r_{0,1}^2 \leq \frac{N_{\text{cell}} - \pi \epsilon r_1^2 \frac{B_{\text{HO}} N_{\text{cell}}}{B_{\text{cell}}}}{\pi \epsilon \big [1 + \beta - \frac{B_{\text{HO}} N_{\text{cell}}}{B_{\text{cell}}} \big]}  :=  \lambda_{0,1}^2. \label{lim_3}
\end{align}
Note that the term $B_{\text{HO}}/B_{\text{sub}}$ in \eqref{lim_1} represents the required number of subcarriers to handle the handover process for each user.
Furthermore, note that $\pi \epsilon \lambda_{0,1}^2$ represents the maximum number of subcarriers that can be allocated in Zone $0$, and hence the maximum number of primary users, for a given $\epsilon$.

\begin{figure}[t]
\begin{center}
\includegraphics[width=0.35\textwidth]{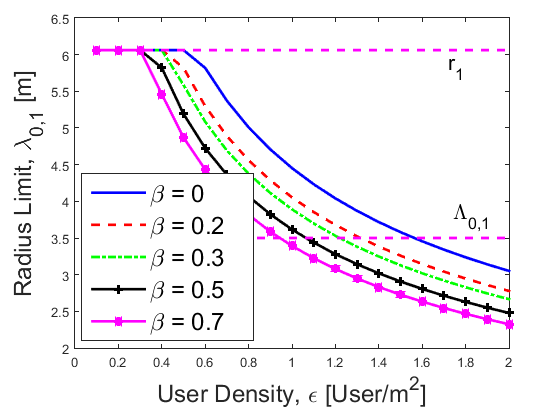}
\caption{Limit on Zone $0$ radius based on the handover requirements as a function of the user density, $\epsilon$, for different values of $\beta$ when $\theta_1 = 60^{\circ}$, $B_{\text{HO}} = 10$ Kbits, and $N_{\text{cell}} = 64$.}\label{fig:r_lim_HO_eps}
\end{center}
\end{figure}

Now, combining \eqref{illumination_3} and \eqref{lim_3}, we can observe that for a given AP, in order to fulfill both illumination and mobility (handover) requirements, while serving at least $U_{\text{pu}}$ primary users for a given user density $\epsilon$, the radius of Zone $0$ should obey the following condition:
\begin{equation}
\label{lim_comb}
\frac{U_{\text{pu}}}{\pi \epsilon} \leq  r_{0,1}^2 \leq \text{min} \{\Lambda_{0,1}^2,\lambda_{0,1}^2\}.
\end{equation}

\mar{Based on the above description, a simple mechanism that supports mobility and handover requirements can be realized in real scenarios as follows~\footnote{Note that developing a handover mechanism is not the main concern of this paper.}. Let a primary user be located in Zone $0$ of a certain lighting cell and be served by that cell (\textit{i.e.} the serving cell). When the user moves out from Zone $0$, it will be switched to another subcarrier of those allocated in Zone $1$ of the serving cell. Note that the switching process within the same cell can be handled without an extra cost, while  if the user enters an overlapping area with an adjacent cell, \textit{i.e.,} Zone $1$ of the adjacent cell, then a handover process can be initiated. During the handover process, the user utilizes some of the resources allocated in Zone $1$ of the candidate cell to handle the switching process, while being connected with the serving cell. The connection with the serving cell is then terminated whenever the handover process is executed.}

In \figurename~\ref{fig:r_lim_HO_eps}, we show the radius limit based on the handover requirements, \textit{i.e.,} $\lambda_{0,1}$, 
as a function of the user density $\epsilon$ and for different values of $\beta$. Herein, we set $\theta_1 = 60^{\circ}$ and $B_{\text{HO}} = 10$ Kbits, while the values of other parameters are as shown in Table~\ref{tab_1}. 
In \figurename~\ref{fig:r_lim_HO_eps} we also show the cell radius (\textit{i.e.,} $r_1$) and 
the radius limit based on the illumination requirements, 
(\textit{i.e.,} $\Lambda_{0,1}$). Here, we set $\lambda_{0,1} = r_1$ if we have $\lambda_{0,1} > r_1$. Note that the value of $\beta = 0$ means that none of the primary users is leaving Zone $0$. This can be the case when all primary users are stationary, or when they (or some of them) are moving within Zone $0$. We initially observe that, at low user density values 
of $\epsilon \leq 0.3$, the handover process has no effects on the cell coverage (flat behavior) for the considered settings 
and values of $\beta$.  This means that the available number of subcarriers can support both the transmission 
and the handover requirements of all users within the entire area of the cell with radius $r_1$. 

As either the user density or the number of 
primary users that move to Zone $1$ increases, \figurename~\ref{fig:r_lim_HO_eps} reveals that the area of 
Zone $0$ shrinks, and hence the number of primary users that can be served by the cell is reduced. 
We finally notice that the handover process has stricter limits on the radius than the illumination requirements 
at higher user density and/or higher values of $\beta$. 
\begin{figure}[t]
\begin{center}
\includegraphics[width=0.35\textwidth]{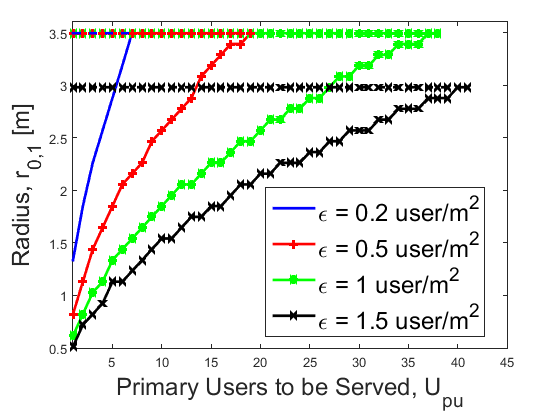}
\caption{Minimum (\textit{solid lines}) and maximum (\textit{dashed lines}) values of radius of Zone $0$ that can satisfy both handover and illumination requirements as a function of the primary users to be served for different values of $\epsilon$, when $\theta_1 = 60^{\circ}$, $B_{\text{HO}} = 10$ Kbits, and $\beta = 0.4$. Herein, 
$\Lambda_{0,1} = 3.5$ m and $\lambda_{0,1} \approx 3$ m.}\label{fig:r_true_Pu}
\end{center}
\end{figure}
In \figurename~\ref{fig:r_true_Pu}, we plot the radius $r_{0,1}$ that satisfies both the handover and illumination requirements, 
as a function of the target number of primary users to be served, \textit{i.e.,} $P_{\text{pu}}$. Here, we show both the minimum (\textit{solid lines}) and maximum (\textit{dashed lines}) values of $r_{0,1}$, \textit{i.e.,} $U_{\text{pu}}/\pi \epsilon$, and $\text{min} \{\Lambda_{0,1},\lambda_{0,1}\}$, respectively, for different values of $\epsilon$, while setting $\beta = 0.4$, $N_{\text{cell}} = 64$, and $\theta_1 = 60^{\circ}$. Note that the minimum values of $r_{0,1}$ in \figurename~\ref{fig:r_true_Pu} also provide the maximum ASE values that can be achieved with the considered parameters and requirements. 

Finally, we observe  that the number of served users is restricted by the illumination constraints at a lower user density, while it is restricted by the mobility (handover) constraints at a higher user density. Note that the maximum number of allocated subcarriers in Zone $0$ for 
$\epsilon = \{0.2,0.5,1,1.5\}$ [user/m$^2$] is $\epsilon \pi \lambda_{0,1}^2 = \{46,45,44,43\}$, respectively.   

\section{Simulation Example}\label{sec:example}
In this section, we investigate the performance of the proposed resource allocation scheme in a practical indoor scenario. In particular, we consider a $10\times30\times3.5$ m$^3$ indoor hall,  covered by $K$ VLC APs with overlapping lighting cells. By assuming that all APs have the same characteristics with a LED half-view angle of 
$\theta = 60^{\circ}$, a number of $K = 3$ APs is  sufficient to cover the entire space. We further assume that the overlapping area between any two adjacent cells has a maximum distance of 
$1.2$~m. 

\begin{figure}[t]
\begin{center}
\includegraphics[width=0.35\textwidth]{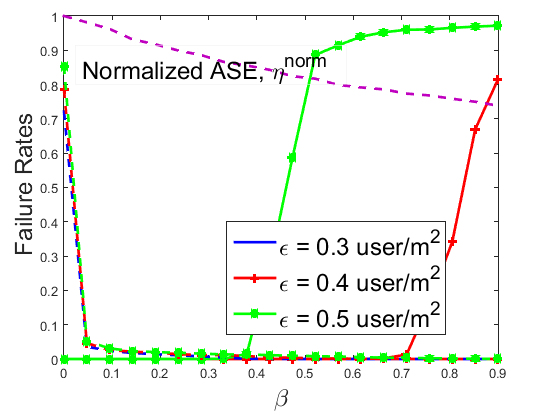}
\caption{Average failure rates $\delta_0$ (\textit{solid lines}) and $\delta_1$ (\textit{dashed lines}) for the simulation example as a function of the ratio $\beta$ for different values of the user density $\epsilon$.  We also show the normalized ASE (\textit{dashed purple line}). }\label{fig:disc_prob}
\end{center}
\end{figure}

The number of users is $M$, and hence the user density can be calculated as $\epsilon = 300/M$ [user/m$^2$]. All users are uniformly distributed and move with random speeds and directions. Particularly, each user moves with a speed uniformly distributed between $0$ and $\nu_{max}$ [m/s]. 
Speed limits are affected by user positions, 
assuming lower (higher) speeds for primary (secondary) users. This is justified since users are expected to be looking for spots with better illumination and communication conditions, which can be found in Zone $0$ of each cell. 
So, we set $\nu_{max} = 0.5$ m/s for users moving in Zone $0$, while 
 $\nu_{max} = 2$ m/s for users in Zone $1$. 
 
 Based on the same criteria, we further assume that the user located in Zone $0$ changes its direction uniformly between $0$ and $2\pi$ in each simulation step, whereas it moves in the direction of the closest AP otherwise. Let $T_{s}$ be the simulation step in seconds, after which the new position of each user is updated. Consequently, each user moves a distance of $T_s \nu$ [m] in each step, where $\nu \sim \mathcal{U}(0,\nu_{max})$ is the user speed. Herein, we set $T_s = 0.5$ s. 
 
In this section, we calculate the radius and the number of allocated subcarriers of Zone $0$ 
related to the $k$-th AP as $\min\{\Lambda_{0,k},\lambda_{0,k}\}$ and $\epsilon \pi \lambda_{0,k}^2$, respectively. Let $U_{z}^{(t)}$ be the total number of users located in Zone $z$ of all APs 
in the scenario at a time instant $t$ for $z = \{0,1\}$. Then, we define the total failure rates in Zone $0$ 
and Zone $1$ over the entire space at the time step $t$, respectively, as the probability that no user is within Zone $0$
and Zone $1$, respectively:
\begin{equation}
 \delta_0^{(t)} = \text{Prob}\{U_{0}^{(t)} > \epsilon \pi \sum_{k=1}^K \lambda_{0,k}^2\}, 
\end{equation}
 and 
\begin{equation}
\delta_1^{(t)}= \text{Prob}\{U_{1}^{(t)} > N_{\text{cell}} - \epsilon \pi \sum_{k=1}^K \lambda_{0,k}^2\}.
\end{equation} 

In \figurename~\ref{fig:disc_prob}, we show the average failure rates, $\delta_0 = \mathbb{E}_t \{\delta_0^{(t)}\}$ and $\delta_1 = \mathbb{E}_t \{\delta_1^{(t)}\}$, over a simulation time of $2$ minutes. 
Here we consider  $K = 3$ with $\theta_k = 60^{\circ}$ for $k = 1,2,3$, $B_{\text{HO}} = 10$ Kbits, and $N_{\text{cell}} = 64$. 
For $\epsilon = \{0.3,0.4,0.5\}$ user/m$^2$, the number of users is $M = \{90,120,150\}$. 
We can clearly see that  
$\beta$ has a potential impact on the  average failure rates for any values of $\epsilon$. As expected, increasing $\beta$ degrades the performance in Zone $0$ in terms of the user connectivity, since less subcarriers are allocated in Zone $0$. On the other hand, increasing $\beta$ results in more subcarriers in Zone $1$, and hence better user connectivity, as also shown in \figurename~\ref{fig:disc_prob}. 

We also plot the normalized ASE factor \textit{i.e.}, $\eta^{\text{norm}}=\eta/\max(\eta)$, for the considered scenario in \figurename~\ref{fig:disc_prob}. Note that the illumination constraints have the main limits on the radius of Zone $0$ for the considered values 
of $\epsilon$, and hence the ASE curve is the same here. While increasing $\beta$ shrinks the area of Zone $0$, it also means that Zone $1$ has a larger area with more allocated subcarriers. Therefore, the ASE performance degrades with $\beta$. 
As a conclusion, optimizing the value of $\beta$ is  a critical task in the proposed scheme, in order to reduce the failure rates in both zones, while achieving the best possible ASE for a certain scenario. 

\section{Conclusion}
In this paper, we have introduced the concept of cognitive VLC networks by distinguishing  primary and secondary users based on their locations. Considering OFDMA-based networks, the regions of primary and secondary users are defined in terms of the radius and the number of allocated subcarriers. Certain conditions on defining the two regions are then derived  to guarantee satisfying illumination, mobility, and handover needs. Simulation results of a real scenario have showed that a proper value for the mobility parameter can fulfill the mentioned requirements, while achieving  high ASE within the cell. 
%
  
\bibliographystyle{IEEEtran}
\bibliography{references}

\end{document}